\documentclass[pra,showkeys,showpacs]{revtex4}

\usepackage{amsmath}
\usepackage{graphicx,epsfig}
\newcommand{\ket}[1]{|#1\rangle}

\begin{document}
\title{Multi-Partite Quantum Cryptographic Protocols with Noisy GHZ States}

\author{Kai Chen}
\author{Hoi-Kwong Lo}

\affiliation{Center for Quantum Information and Quantum Control\\
Department of Physics and Department of Electrical \& Computer Engineering\\
University of Toronto, Toronto, Ontario, M5S 3G4, Canada}

%\date{Received November 30, 2006; revised May 14, 2007}

\begin{abstract}
We propose a wide class of distillation schemes for multi-partite
entangled states that are CSS-states. Our proposal provides not
only superior efficiency, but also new insights on the connection
between CSS-states and bipartite graph states. We then apply our
distillation schemes to the tri-partite case for three
cryptographic tasks---namely, (a) conference key agreement, (b)
quantum sharing of classical secrets and (c) third-man
cryptography. Moreover, we construct ``prepare-and-measure''
protocols for the above three cryptographic tasks which can be
implemented with the generation of only a single entangled pair
at a time. This gives significant simplification over previous
experimental implementations which require two entangled pairs
generated simultaneously. We also study the yields of those
protocols and the threshold values of the fidelity above which
the protocols can function securely. Rather surprisingly, our
protocols will function securely even when the initial state does
not violate the standard Bell-inequalities for GHZ states.
\end{abstract}



\maketitle

\section{Introduction}
\noindent
\subsection{Motivations}
\noindent
Entanglement is the hallmark of quantum mechanics.
There are many different types of entanglement. While the
classification of bipartite pure-state entanglement has been
solved, currently the classification of multi-partite
entanglement and mixed state bipartite entanglement is a most
important open problem (for a recent review, see
\cite{Horo-qic}). A particularly useful idea for the study and
quantification of entanglement is entanglement distillation
\cite{Betal,BDSW}). Given a large number, say $n$, of copies of
an initial mixed state $\rho$, one may ask how many, $m$,
standard states (e.g., Einstein-Podolsky-Rosen (EPR) states and
Greenberger-Horne-Zeilinger (GHZ) states) one can distill out in
the end by using only local operations and classical
communications? The ratio $m/n$, in the limit of large $n$,
optimized over all possible procedures denotes the amount of
distillable entanglement in the original state, $\rho$.
Unfortunately, except for the case of bipartite pure states, it
is generally hard to work out the optimal strategy for
entanglement distillation. A simpler problem is to construct
explicit strategies for entanglement distillation. Their yields
will give lower bounds on the amount of distillable entanglement.

Recently, the distillation of entanglement from multi-partite
entangled states has attracted a lot of attention
\cite{mppvk98,dctprl99,dcpra00,smolincm2002,graph,graph1}. Direct
distillations of multipartite entanglement (such as GHZ states)
have some advantages over concatenated distillation of partite
entanglement (EPR states). Indeed, for the GHZ state, direct
distillations have been shown to be more efficient than
concatenated distillation of partite entanglement
\cite{mppvk98,smolincm2002,graph,graph1}. Moreover, they tolerate
a higher noise level \cite{mppvk98,smolincm2002,graph,graph1}.

Another interesting line of research is the cryptographic
applications of multi-partite entanglement. In particular, the
idea of quantum sharing of classical secrets has been proposed in
\cite{hbb99}. Such protocols guarantee security against an
eavesdropper.

In this paper, we study the distillation of {\it multi}-partite
entanglement and its applications to multi-party quantum
cryptography. Our work provides the missing links between a
diverse range of protocols and breaks several new grounds. First,
whereas most previous results refer to the distillation of EPR
states and GHZ states only, our formalism applies to a much wider
class of states---CSS states. See Subsection~\ref{ss:CSSstates}
for a definition.

Second, we show that CSS-states are mathematically equivalent to
bi-partite graph states. This insight allows one to reformulate
the previous work on the distillation of bi-partite graph states
\cite{graph,graph1} in the systematic CSS formalism. Note that
much of \cite{graph,graph1} dealt with recurrence protocols whose
yields go to zero in the limit that the fidelity of the final
states tend to 1. In contrast, our re-formulation allows the
direct application of hashing type protocols that give non-zero
yields in the same limit.

Third, we construct distillation protocols of the GHZ state which
give higher yields and tolerate much higher levels of noises than
previous protocols. Our improved protocol exploits the
non-trivial amount of mutual information between various
variables which characterize a state, thus highlighting the power
of systematic applications of concepts in classical information
theory to entanglement distillation.

Fourth, we construct ``prepare-and-measure'' type protocols for
multi-partite quantum cryptography. In a prepare-and-measure
protocol, a (potentially dishonest) preparer sends an ordered
sequence of identical multi-party entangled states to multiple
parties through some noisy channels. The participants then
perform some {\it local individual} quantum measurements and
local {\it classical} computations and classical communications
(CCCCs) between them. A prepare-and-measure type protocol is of
practical interest because it has an advantage of being
technologically less demanding. To implement a
prepare-and-measure protocol, the participants do not need to
have full-blown quantum computers. The quantum computational part
is done solely by a preparer (who prepares some standard
entangled multi-partite state). What the participants need to do
is to perform some local individual measurements. The rest are
CCCCs that can be performed by strictly classical devices.

We note that the structure of a distillation protocol is directly
relevant to its reduction here. In fact, in this paper, we find
that a sufficient condition for a distillation protocol to be
reducible to a prepare-and-measure type protocol is that it is
distilling a CSS state and that the procedure involves
post-selection steps in at most one of the two types (X-type or
Z-type) of observables. This criterion generalizes the finding of
Gottesman-Lo \cite{GLIEEE03} to the multi-partite case. [Indeed,
our study of prepare-and-measure type protocol for multi-party
quantum cryptography is a generalization of the security proofs
of BB84 by Shor-Preskill \cite{Shor-Preskill} and Gottesman-Lo
\cite{GLIEEE03}. The first proof of security of BB84 was by
Mayers \cite{mayersqkd}.]

We note that an $N$-party prepare-and-measure quantum
cryptographic protocol only requires an $(N-1)$-partite
entanglement for its implementation. This is because the preparer
has the option to measure her state immediately.

Fifth, we propose new protocols in multi-partite quantum
cryptography. Specifically, we propose a conference key agreement
protocol for three parties. Such a protocol can also be used to
implement a previous protocol---third man cryptography.

Sixth, we give quantitative number to the yields of our
protocols. Moreover, our protocols can tolerate rather high error
rates. Surprisingly, our protocols will function securely even
when the initial state fails to violate the standard
Bell-inequalities for GHZ states. This observation may be of
interest to the foundations of quantum theory.

Given that experimental works on quantum sharing of classical
secrets and third-man quantum cryptography have recently been
done \cite{expQSCS}, our work is very timely for their security
analysis.

Seventh, we propose a simpler experimental realization of quantum
sharing of classical secret schemes and third-man cryptography.
Whereas previous experimental implementations \cite{expQSCS}
required the challenging feat of the generation of two entangled
pairs via parametric down conversion, our new proposal has a
clear advantage in requiring the generation of only a single EPR
pair at a time. This leads not only to a substantial
simplification of the experiment, but also a much higher rate.
Our simpler protocols are based on the following idea. For the
case of three parties, we convert our protocols to those
involving only {\it bi}-partite entanglement. Therefore, our
protocols have clear near term applications.

\subsection{Organization of the paper}
\noindent
Our paper is organized as follows. In Section~2, we study the
entanglement distillation of the GHZ state and present an
improved hashing protocol. In Section~3, we generalize our
results from the GHZ state to a general CSS state and show that
the various subroutines that we have studied, in fact, apply to a
general CSS state. In Section~4, we show the equivalence of CSS
states and bipartite graph states, thus connecting our work to
other works \cite{graph,graph1}. One part of this equivalence is
due to Eric Rains \cite{rains}. In section~5, we discuss
multi-party quantum cryptography and our adversarial model. In
Section~6, we apply our formulation to study the three-party
conference key agreement problem and show that for tripartite
Werner state, conference key agreement is possible whenever the
fidelity $F\geq  0.3976$. We note that our three-party conference
key agreement protocol can be implemented with only bi-partite
entanglement. In Section~7, we study the secret sharing problem
and show that Alice can successfully share a secret with two
parties Bob and Charlie if they share a Werner state with
fidelity $F \geq 0.5372$. As before, our protocol can be
implemented by using bi-partite entanglement only. We demonstrate
in Section~7, that our protocols, which involve two-way classical
communications, are provably better than any protocols involving
only one-way classical communications. Finally, we show that,
rather surprisingly, our protocols will work even when the
initial GHZ-state does not violate the standard Bell inequalities
for a GHZ-state.

\section{Distillation of the GHZ state}
\noindent
\subsection{Some notations}
\noindent
Suppose three distant parties, Alice, Bob, and Charlie, share
some tri-partite state. An example of a well-known tri-partite
pure state is the GHZ state. It has the form of
\begin{equation}
\left\vert \Psi \right\rangle
_{ABC}=\frac{1}{\sqrt{2}}(\left\vert 0\right\rangle \left\vert
0\right\rangle \left\vert 0\right\rangle +\left\vert
1\right\rangle \left\vert 1\right\rangle \left\vert
1\right\rangle) .
\end{equation}
The GHZ state is the $+1$ eigenstate of the following set of
commuting observables
\begin{eqnarray}
S_{0}& =&X\otimes X\otimes X, \nonumber \\
S_{1}& =&Z\otimes Z\otimes I, \nonumber \\
S_{2}& =&Z\otimes I\otimes Z, \label{e:stabilizer}
\end{eqnarray}
where $X=$ $%
\begin{pmatrix}
0 & 1 \\
1 & 0%
\end{pmatrix}%
,Y=%
\begin{pmatrix}
0 & -i \\
i & 0%
\end{pmatrix}%
,Z=%
\begin{pmatrix}
1 & 0 \\
0 & -1%
\end{pmatrix}$.
In the stabilizer formulation, the three observables are the
stabilizer generators of the GHZ state. For simplicity, we denote
them by $XXX$, $ZZI$ and $ZIZ$. By multiplying a combination of
them together, we obtain other non-trivial stabilizer elements
\begin{equation}
-YYX,IZZ,-YXY,-XYY.
\end{equation}

Let us define a GHZ basis as a three-qubit basis whose basis
vectors are eigenvectors of the stabilizer generators listed in
(\ref{e:stabilizer}). More concretely, its basis vectors are
\begin{equation}
\left\vert \Psi_{p,i_{1},i_{2}}\right\rangle _{ABC}=\frac{1}{\sqrt{2}}%
(\left\vert 0\right\rangle \left\vert i_{1}\right\rangle
\left\vert
i_{2}\right\rangle +(-1)^{p}\left\vert 1\right\rangle \left\vert \overline{%
i_{1}}\right\rangle \left\vert \overline{i_{2}}\right\rangle ,
\label{GHZbasis}
\end{equation}
where $p$ and the $i$'s are zero or one and a bar over a bit
value indicates its logical negation. The $p$ corresponds to
whether a state is a +1 or -1 eigenvector of $S_0$ which we call
the ``phase¡± bit of the state Eq.~\ref{GHZbasis}. The $i_j$
correspond to whether a state is a +1 or -1 eigenvector of $S_j$
for j = 1,2, which we call the amplitude bits. More specifically,
the three labels $(p,i_{1},i_{2})$ correspond to the eigenvalues
of the 3 stabilizer generators $S_{0},S_{1},S_{2}$ by
correspondence relation
\begin{align*}
\text{eigenvalue \ \ }1 & \longrightarrow\text{label }0, \\
\text{eigenvalue}-1 & \longrightarrow\text{label }1.
\end{align*}

The GHZ-basis has $8$ basis vectors and they may be labelled by
$8$ alphabets, $\alpha$, $\beta$, $\gamma$, $\delta$, etc.
Thus, we can label a density matrix which is diagonal in the
GHZ-basis by the following form
\begin{eqnarray}
\rho_{ABC}= \left(
\begin{array}{cccccccc}
p_{000} & 0 & 0 & 0 & 0 & 0 & 0 & 0 \\
0 & p_{100} & 0 & 0 & 0 & 0 & 0 & 0 \\
0 & 0 & p_{011} & 0 & 0 & 0 & 0 & 0 \\
0 & 0 & 0 & p_{111} & 0 & 0 & 0 & 0 \\
0 & 0 & 0 & 0 & p_{010} & 0 & 0 & 0 \\
0 & 0 & 0 & 0 & 0 & p_{110} & 0 & 0 \\
0 & 0 & 0 & 0 & 0 & 0 & p_{001} & 0 \\
0 & 0 & 0 & 0 & 0 & 0 & 0 & p_{101}
\end{array}
\right) .
\nonumber \\
\label{e:GHZdiagonal}
\end{eqnarray}

Similarly, given $n$ trios of qubits, we define the $n$-GHZ-basis
whose basis vectors are the tensor product states of the
individual GHZ-basis states. An $n$-GHZ-basis state can be
labelled by a $n$-string of the $8$ alphabets ($\alpha$, $\beta$,
etc).

\subsection{Non-identical independent distribution}
\noindent
In this paper, we often assume three distant parties, Alice, Bob
and Charlie, share $n$ trios of qubits. Notice that there is no
need to assume an i.i.d. (independent identical distribution) for
the $n$ trios. Instead, we consider the most general setting
where those $n$ trios can be fully entangled among themselves and
perhaps also with some additional ancillas held by an
eavesdropper, Eve. In subsequent paragraphs, we will describe a
depolarizing procedure, which can be performed by local
operations and classical communications (LOCCs) by Alice, Bob and
Charlie, that will turn the state of the $n$ trios to one that is
diagonal in the $n$-GHZ-basis mentioned in the last paragraph.

We will then describe an estimation procedure that will allow one
to estimate the {\it type} of the $n$-string. [That is, to say
the relative frequencies of those $8$ alphabets in the string.]
After that, we will introduce some entanglement distillation
protocols (based on recurrence protocol and hashing protocols)
that will work well, depending only on the type of the string.

Since in all the aforementioned procedures the important point is
the type of a string, in what follows, it suffices in our
analysis to consider the marginal density matrix (of the
depolarized state) of a single trio. This is because this
marginal density matrix precisely captures all the information
about the type of a string.

\subsection{Depolarization to the GHZ-basis diagonal states}
\noindent
Suppose three parties share a general tri-partite density matrix
that is not necessarily diagonal in the GHZ-basis. In this
subsection, we recall \cite{dctprl99,dcpra00} a general (LOCC)
procedure that will allow the three parties to depolarize the
state to one that is diagonal in the GHZ-basis by the following
steps. Start with the operator $XXX$, the three parties apply it
with a probability $1/2$. Note that $XXX$ is a tensor product of
local unitary transformations and as such can be applied by
LOCCs. Now, consider the next operator, $ZZI$, the three parties
apply it with a probability $1/2$. Finally, consider the operator
$ZIZ$, they also apply it with a probability $1/2$. The overall
operation corresponds to
\begin{eqnarray}
\rho &\longrightarrow &
\frac{1}{8}\Big(\rho +(XXX)\rho (XXX)+(ZZI)\rho (ZZI)
 \nonumber \\
&& +(YYX)\rho (YYX)+(ZIZ)\rho (ZIZ)+(YXY)\rho (YXY)
 \nonumber \\
&& +(IZZ)\rho (IZZ)+(ZYY)\rho (ZYY)\Big).
\end{eqnarray}
The overall operation makes $\rho $ diagonal in the basis of
states in Eq.~(\ref{GHZbasis}) without changing the diagonal
coefficients. If one prefers, one can also arrange that
$p_{011}=p_{111},p_{010}=p_{110}$ and $p_{001}=p_{101}$ by random
operations
$\begin{pmatrix}
e^{i\phi _{\alpha }} & 0 \\
0 & 1
\end{pmatrix}$
(up to a factor of any combination of the eight group elements, e.g.
$X\begin{pmatrix}
e^{i\phi _{\alpha }} & 0 \\
0 & 1%
\end{pmatrix}%
=%
\begin{pmatrix}
0 & 1 \\
e^{i\phi _{\alpha }} & 0%
\end{pmatrix}$
does the same thing) where $\alpha =A,B,C$ and satisfies $\phi
_{A}+\phi _{B}+\phi _{C}=0.$ See \cite{dctprl99,dcpra00} for
details.

Now, given $n$ trios of qubits, Alice, Bob and Charlie can apply
the above randomization process to each trio. Therefore, they can
turn the state of the $n$ trios into a $n$-GHZ-basis diagonal
state.

\subsection{Estimation of relative frequencies of eight alphabets}
\noindent
Now, Alice, Bob and Charlie are interested in estimating the type
of the $n$-string in the $n$-GHZ-basis. This can be achieved by
the following random sampling argument. [This is related to the
commuting observable argument in \cite{qkd}.]

More concretely, if Alice, Bob and Charlie pick a random sample
of $m$ out of the $n$ trios and for each trio, measure along $X,
Y, Z$ basis (i.e., measure $X$, $Y$ and $Z$ observables) and
compare the results of their local measurements, they can
estimate the diagonal matrix elements in (\ref{e:GHZdiagonal}).
In the limit of an infinite ensemble, they will find that the
error rate for the seven non-trivial group elements is
\begin{align}
XXX & :p_{100}+p_{101}+p_{110}+p_{111}, \nonumber \\
ZZI & :p_{010}+p_{011}+p_{110}+p_{111}, \nonumber \\
ZIZ & :p_{001}+p_{011}+p_{101}+p_{111}, \nonumber \\
-YYX & :p_{100}+p_{011}+p_{010}+p_{101}, \nonumber \\
IZZ & :p_{010}+p_{110}+p_{001}+p_{101}, \nonumber \\
-YZY & :p_{100}+p_{011}+p_{110}+p_{001}, \nonumber \\
-XYY & :p_{100}+p_{111}+p_{010}+p_{001}.
\end{align}

\vspace{0.3cm} \noindent \textbf{Remark 1:} Here, by an error
rate of an observable, we mean the probability of getting a $-1$
eigenvalue. Note that an error occurs for an observable
$-YYX=XXX\times ZZI,$ if either $XXX$ or $ZZI$ has an error, but
not both.

\vspace{0.3cm} \noindent \textbf{Remark 2:} If denoting the error
rates for all of the 7 non-trivial group elements by $s_1,
\ldots, s_7$, one has
\begin{align}
p_{000}& =1-(s_{1}+s_{2}+s_{3}+s_{4}+s_{5}+s_{6}+s_{7})/4, \nonumber \\
p_{100}& =(s_{1}-s_{2}-s_{3}+s_{4}-s_{5}+s_{6}+s_{7})/4, \nonumber \\
p_{011}& =(-s_{1}+s_{2}+s_{3}+s_{4}-s_{5}+s_{6}-s_{7})/4, \nonumber \\
p_{111}& =(s_{1}+s_{2}+s_{3}-s_{4}-s_{5}-s_{6}+s_{7})/4, \nonumber \\
p_{010}& =(-s_{1}+s_{2}-s_{3}+s_{4}+s_{5}-s_{6}+s_{7})/4, \nonumber \\
p_{110}& =(s_{1}+s_{2}-s_{3}-s_{4}+s_{5}+s_{6}-s_{7})/4, \nonumber \\
p_{001}& =(-s_{1}-s_{2}+s_{3}-s_{4}+s_{5}+s_{6}+s_{7})/4, \nonumber \\
p_{101}& =(s_{1}-s_{2}+s_{3}+s_{4}+s_{5}-s_{6}-s_{7})/4.
\end{align}
Since $s_1, \ldots, s_7$ can be determined by LOCCs by Alice, Bob
and Charlie, the above equations relate the diagonal matrix
element of the marginal density matrix $\rho_{ABC}$ of the
measured trios to experimental observables.

\vspace{0.3cm} \noindent \textbf{Remark 3:} In practice, only $m$
out of the $n$ trios are used as test trios. Nonetheless, the
relative frequencies of the eight alphabets in the $m$ test trios
provide a reliable estimate of the relative frequency of the
remaining $n-m$ trios. Bounds on random sampling estimation
procedure has been discussed in the context of EPR pairs in
\cite{Shor-Preskill,Gottesman-Preskill,eff}. The estimation
procedure for a tri-partite (GHZ) state presented in this
subsection is a natural generalization of the estimation
procedure of a bi-partite (EPR) state presented in
\cite{losixstate01}, in which the mutual information between
variables is taken into account.

\subsection{Multi-party hashing for
distillation of GHZ states}
\noindent
We recall the results of Maneva and Smolin \cite{smolincm2002} on
multi-partite entanglement distillation. Suppose $N$ ($> 2$)
parties share $n$ (generally non-i.i.d.) mixed multi-partite
states and they would like to distill out almost perfect
(generalized) GHZ states. By using the notation in
\cite{smolincm2002}, one can write an unknown $N$-qubit state as
an $N$-bit string $p,i_{1},i_{2},\ldots ,i_{N-1}$ where $p$
corresponds to the eigenvalue of the operator $XX \cdots X$ and
$i_j$ ($j \geq 1$) corresponds to the eigenvalue of the operator
$Z_1 Z_{j+1}$. (See Eq.~(3) of \cite{smolincm2002}.) Note that
$p$ denotes the phase error pattern (error syndrome) and $i_j$
($j \geq 1$) denotes the bit-flip error pattern.

\begin{figure} [htbp]
%\vspace*{13pt}
\centerline{\epsfig{file=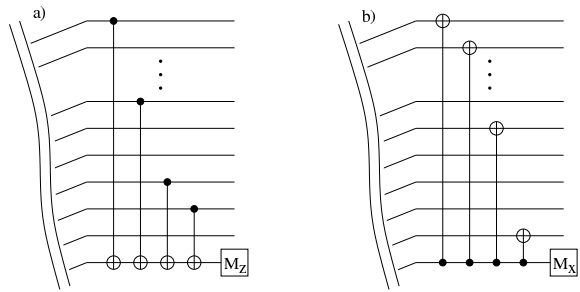, width=8.2cm}} %100 percent
\vspace*{13pt}
\caption{\label{fig1}Multi-party hashing circuits of Maneva and Smolin.
Re-produced from Fig.~4 of \cite{smolincm2002}. ``These hashes
are done on large blocks of bits (indicated by the vertical
ellipsis) and are done multilaterally (only one party's
operations are shown, the other $N-1$ parties operations are
identical). a) Finding a random subset parity on all the
$b_{j>0}$ in parallel. In this case, the first, third, sixth and
seventh states shown are XORed multilaterally into the last one
which is then measured to determine the eigenvalue of the $Z$
operator. b) Finding a random subset parity on $b_0$. In this
instance the parity of the first, second, fourth and eighth
states shown are XORed with the last one, which is then measured
in the eigenbasis of the $X$ operator. Note the reversal of the
direction of the XOR gates with respect to a)." }
\end{figure}

Maneva and Smolin \cite{smolincm2002} found a high-yield
multi-partite entanglement distillation protocol---multi-party
hashing method. Their protocol generalizes the hashing protocol
in the bipartite case \cite{BDSW}. Maneva and Smolin used
multilateral quantum XOR gates (MXOR) as shown in the
Fig.~\ref{fig1}. The basic property of the MXOR gate acting on
two GHZ-like states is
\begin{eqnarray}
&&\text{MXOR}[(p,i_{1},i_{2},\ldots
,i_{N-1}),(q,j_{1},j_{2},\ldots
,j_{N-1})] \nonumber \\
&&= [(p\oplus q,i_{1},i_{2},\ldots ,i_{N-1}), (q,i_{1}\oplus j_{1},i_{2}\oplus j_{2},\ldots
,i_{N-1}\oplus j_{N-1})],
\end{eqnarray}
where $(p,i_{1},i_{2},\ldots ,i_{N-1})$ and
$(q,j_{1},j_{2},\ldots ,j_{N-1})$ denote the phase bit and
amplitude bits for the first and second GHZ-like states.

Consider large number $n$ (generally non-i.i.d.) multi-partite
states. Let $\hat{B_0}$ denote a vector of $n$ values that
represents the random variable $p$'s for the $n$ multi-partite
states, while each of $\hat{B_1} \hat{B_2} \cdots \hat{B_{N-1}}$
a vector of $n$ values for random variable $i_{1},i_{2},\cdots
,i_{N-1}$ respectively for the $n$ multi-partite states. Let
$b_0$ denotes a random variable that represents the result of a
random choice of an element in $\hat{B_0}$. Similarly, let $b_1,
b_2, \cdots ,  b_{N-1}$ respectively denote random variables that
represent the result of random choices of $\hat{B_1} \hat{B_2}
\cdots \hat{B_{N-1}}$. In other words, $H(b_0)$ denotes the
averaged phase error rate over the $n$ multi-partite states
whereas $H(b_i)$ ( $ 1 \leq i \leq N-1)$ denotes the averaged
bit-flip error over the $n$ multi-partite states for the $i$-th
Z-type stabilizer generator.

Let us recall the following fact in random hashing. {\it
Fact~One}: If a vector is drawn from a set of size $2^{s}$ and a
random linear hash is obtained with $s + O (\log ( 1 /
\epsilon))$ bits of output, then one can identify the vector
uniquely with a probability at least $ 1 - \epsilon$.

In this paper, we are interested in the asymptotic limit of large
$n$. Maneva and Smolin showed that the asymptotic yield (per
input mixed state) of their hashing protocol is given by
\begin{equation}
D_{h}=1-\max_{j>0}[\{H(b_{j})\}]-H(b_{0}),  \label{smolinhash}
\end{equation}
where $H(x)$ is the standard Shannon entropy in classical
information theory \cite{Cover}. ($H(x) = - \sum_i p_i \log_2
p_i$ where $p_i$'s are the probabilities for the various distinct
outcomes of a random variable $x$.) We define $H(b_{j})$ $(j=0,
1, \cdots, N-1)$ here to be the entropy per bit in string $b_j$.

Maneva and Smolin's protocol consists of two steps. First, the
parties perform a number of rounds of random hashing in the
amplitude bits using the circuit shown in Fig.~\ref{fig1}a).
Second, they perform a number of rounds of random hashing in the
phase bit(s) using the circuit shown in Fig.~\ref{fig1}b).

Consider the first step---random hashing in amplitude bits. Here
the goal is to find out the identity of the amplitude bits,
$b_j$'s. Note that in each round of hashing of the amplitude
bits, one {\it simultaneously} obtains multiple hash values, one
for each variable $b_j$. Maneva and Smolin argued that the worst
case scenario occurs when the variables, $b_{j_1}$ and $b_{j_2}$,
are independent.

Recall the aforementioned Fact~One in (classical) random hashing
that each round of hashing essentially reduces the entropy of a
string by one bit. The number of rounds needed for amplitude
hashing will, therefore, be given by $n
\max_{j>0}[\{H(b_{j})\}]$. [To be more precise, we mean $n \left(
\max_{j>0}[\{H(b_{j})\}] + \delta \right)$.] Now consider the
second step---random hashing in phase bit(s). In the case of GHZ
state, there is only one phase bit and, therefore, $n H(b_0)$
rounds of random hashing is needed to find out its value.

Let us now present our improvement over Maneva and Smolin. In
what follows, we will argue that, in fact, the yield can be
increased to be
\begin{equation}
D_{h}^{^{\prime }}=1-\max
\{H(b_{1}),H(b_{2}|b_{1})\}-H(b_{0})+I(b_{0};b_{1},b_{2}),
\label{newyield}
\end{equation}
where $I(X;Y)$ is the standard mutual information between two
random variables in classical information theory \cite{Cover}.
Note that there may be some correlations between $b_{0}$ and
$(b_{1},b_{2})$ and also between $b_{1}$ and $b_{2}$ i.e.,
\begin{eqnarray}
I(b_{0};b_{1},b_{2}) &\geq &0, \nonumber \\
I(b_{1};b_{2}) &\geq &0.
\end{eqnarray}
If these quantities are non-zero, we now argue that the parties
can reduce the number of rounds of hashing relative to
Maneva-Smolin's protocol. Consider the following strategy
\begin{enumerate}
\item[1] Alice, Bob and Charlie apply random hashing according to
Fig.~\ref{fig1}a to identify the pattern of $b_{1},b_{2}$. We now
argue that only slightly more than $n [\max
\{H(b_{1}),H(b_{2}|b_{1})\}]$ rounds of random hashing is needed
(we suppose here $H(b_{1})\leq H(b_{2})$). This is because of the
following. We can imagine that the parties perform $n H(b_{1})$
rounds of hashing to work out the value of $b_1$'s completely.
Afterwards, the uncertainty in the variables $b_{2}$'s is reduced
to $n H(b_{2}|b_{1})\}$. Therefore, only $n H(b_{2}|b_{1})\}$
rounds of hashing is needed to work out the value of $b_2$'s.
Note that the hashing of $b_1$ and $b_2$ can be simultaneously
executed. Therefore, in total, we still only need $n [\max
\{H(b_{1}),H(b_{2}|b_{1})\}]$ rounds of random hashing in
amplitude bits.

\item[2] They use the information on the pattern of $b_{1},b_{2}$
(the amplitude bits) to reduce their ignorance on the pattern of
$b_{0}$ (the phase bit) from $n H(b_{0})$ to $n
H(b_{0}|(b_{1},b_{2}))=$ $n [ H(b_{0})-I(b_{0};b_{1},b_{2})] =n [
H(b_{0},b_{1},b_{2})-H(b_{1},b_{2})].$

\item[3] They apply a random hashing through Fig.~\ref{fig1}b to
identify the pattern
of $b_{0}$. Now only (slightly more than)
$n [H(b_{0},b_{1},b_{2})-H(b_{1},b_{2})] $ rounds of random
hashing is needed.
\end{enumerate}

The yield of our method gives
\begin{equation}
D_{h}^{^{\prime }}=1-\max
\{H(b_{1}),H(b_{2}|b_{1})\}-H(b_{0})+I(b_{0};b_{1},b_{2}).
\label{ourhashyield}
\end{equation}

\subsection{Werner-like states}
\noindent
Werner-like states are a generalization of the Werner state
\cite{Werner89} to the multi-partite case and has the form
\begin{equation}
\rho _{W}=\alpha \left\vert \Phi ^{+} \right\rangle \langle \Phi ^{+} |+\frac{%
1-\alpha }{2^{N}}I,\text{ \ }0\leq \alpha \leq 1,
\label{Werner}
\end{equation}
where $\ket{\Phi ^{+}} $ denotes the so-called cat state (i.e.,
GHZ-state for the tri-partite case) and $I$ is the identity
matrix. The fidelity $\rho _{W}$ is $F=\langle \Phi ^{+} |\rho
_{W}\left\vert \Phi ^{+} \right\rangle =\alpha +\frac{1-\alpha
}{2^{N}}$. Using the GHZ-basis, we can rewrite it into
\begin{eqnarray}
\rho _{W}=F\left\vert 0,00\ldots 0\right\rangle \langle 0,00\ldots 0|
+\frac{1-F}{2^{N}-1}(I-\left\vert 0,00\ldots 0\right\rangle
\langle 0,00\ldots 0|).
\end{eqnarray}

\begin{figure} [htbp]
%\vspace*{13pt}
\centerline{\epsfig{file=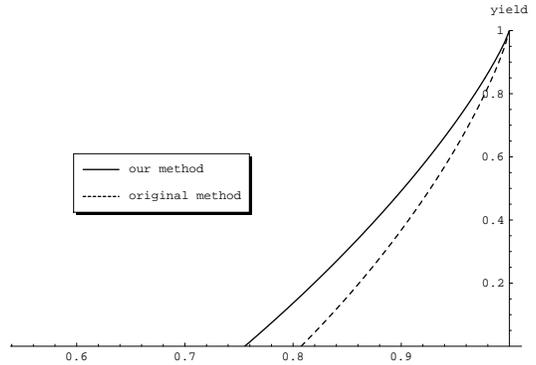, width=8.2cm}} %100 percent
\vspace*{13pt}
\caption{\label{fig2}Comparison of our improved hashing method with the
original method of Maneva and Smolin's \cite{smolincm2002} for a
Werner-like state with initial fidelity $f$ (see
(\ref{Werner})).}
\end{figure}

\vspace{0.3cm} \noindent \textbf{Remark 4:} Using the random
hashing method of Maneva and Smolin, for the tri-partite case,
one obtains perfect GHZ states with nonzero yield whenever $F\geq
0.8075$ from Eq.~(\ref{smolinhash}). With our improved random
hashing method in Eq.~(\ref{ourhashyield}), we get a nonzero
yield whenever $F\geq 0.7554$. This is a substantial improvement
of the original method. We show the yield in Fig.~\ref{fig2}.

\subsection{Comparison of two random hashing methods}
\noindent
We show that, for the tri-partite case, our method gives a
substantial improvement over the original method by two specific
examples.

\textit{Example 1:} Suppose we have a Werner-like state with
initial fidelity of $F=0.9$. Direct calculation gives
$H(b_{0})=H(b_{1})=H(b_{2})=0.316$, $I(b_{1};b_{2})=0.074$,
$I(b_{0};b_{1},b_{2})=0.124$. We have the yield
\begin{eqnarray*}
D_{h} &=&1-\max \{H(b_{1}),H(b_{2})\}-H(b_{0}) \\
&=&1-H(b_{1})-H(b_{0}) \\
&=&0.368 ,
\end{eqnarray*}
through the original method while our improved method gives
\begin{eqnarray*}
D_{h}^{^{\prime }} &=&1-\max
\{H(b_{1}),H(b_{2}|b_{1})\}-H(b_{0})+I(b_{0};b_{1},b_{2}) \\
&=&1-H(b_{1})-H(b_{0})+I(b_{0};b_{1},b_{2}) \\
&=&0.492 .
\end{eqnarray*}
We see that one can obtain a substantially higher yield with our
method. Our method
reduces the number of rounds for identifying the pattern of phase bit (from
$H(b_{0})=0.316$ in the original method to
$H(b_{0}|(b_{1},b_{2}))=0.192$). Though here $I(b_{1};b_{2})>0$,
one still needs $n H(b_{1})$ rounds of random hashing for
identifying the pattern of amplitude bit.

\textit{Example 2:} Suppose we have a GHZ-basis diagonal state
with initial fidelity of $F=0.9$ and the diagonal elements for
its density matrix are
\begin{equation*}
\left(
\begin{array}{c}
p_{000} \\
p_{100} \\
p_{011} \\
p_{111} \\
p_{010} \\
p_{110} \\
p_{001} \\
p_{101}
\end{array}
\right) =\left(
\begin{array}{c}
0.9 \\
0.01 \\
0.01 \\
0.01 \\
0.015 \\
0.015 \\
0.02 \\
0.02
\end{array}
\right) .
\end{equation*}
Direct calculation gives $H(b_{0})=0.307$, $H(b_{1})=0.286$,
$H(b_{2})=0.328$, $I(b_{1};b_{2})=0.040$ and
$I(b_{0};b_{1},b_{2})=0.138$. Thus we have the yield
\begin{equation*}
D_{h}=0.365,
\end{equation*}
through the original method while our improved method gives
\begin{equation*}
D_{h}^{^{\prime }}=0.543.
\end{equation*}
For this state, our method reduces not only the number of rounds
for identifying the pattern of amplitude bits (from
$H(b_{2})=0.328$ in the original method to
$H(b_{2}|b_{1})=0.288$), but also the number of rounds for
identifying the pattern of phase bit (from $H(b_{0})=0.307$ in
the original method to $H(b_{0}|(b_{1},b_{2}))=0.169$).

\subsection{Recurrence method for GHZ state distillation (Murao et al's method)}
\noindent
Hashing protocols are particularly useful at low noise level
because they have rather high yields. However, hashing protocols
generally can tolerate rather low noise levels. That is to say,
the threshold value of the fidelity above which hashing protocols
will work is rather high. Fortunately, another class of
protocols---recurrence protocols---enjoys a lower threshold value
for fidelity.

\begin{figure} [htbp]
%\vspace*{13pt}
\centerline{\epsfig{file=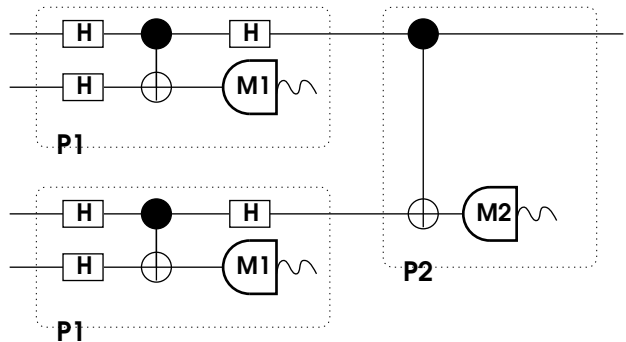, width=8.2cm}} %100 percent
\vspace*{13pt}
\caption{\label{fig3}Purification protocol P1+P2 used in Murao et al's paper
\cite{mppvk98}.}
\end{figure}

We will now describe recurrence protocols. Our idea works for a
general $N$-partite case. However, for concreteness, we will
focus on the tri-partite case. In Murao et al's paper
\cite{mppvk98}, two steps called P1 and P2 are used alternately.
The P1 and P2 steps are shown in Fig.~\ref{fig3}. In P1, the
three parties first randomly permute their trios of qubits.
Consider now a random pair of trios of qubits shared by Alice,
Bob and Charlie. Each party has a pair of qubits in his/her hand.
He/she measures $X_1 X_2$ and broadcasts the outcome. They then
keep the first trio of qubits if there are an even number of $-1$
outcomes. Otherwise, they throw away the two trios of qubits
altogether. Similarly, in P2, each party takes a pair of qubits
as the input and measures $Z_1 Z_2$ and broadcasts the outcome.
They then keep the first trio of qubits if all parties get the
same measurement outcome. Otherwise, they throw away the two
trios of qubits altogether.

We remark that the subroutine P1 checks the parity check
condition, $XXX= + 1$, and the subroutine P2 checks the parity
check conditions, $ZZI= +1 $ and $ZIZ =+1$.

They showed that one can obtain perfect GHZ states by P1+P2
iteratively whenever $F\geq 0.4073$. In fact, one can obtain
better bounds to give $F\geq 0.3483$ by applying P2+P1
iteratively. It is proved in \cite{dcpra00} that one can distill
GHZ state whenever $F>0.3$ by applying a different distillation
protocol. Also D\"ur and Cirac verified in \cite{dcpra00}
numerically that one can also distill GHZ state whenever $F>0.3$
by a modified version of Murao et al's protocol \cite{mppvk98},
which involves a state-dependent sequence of P1 and P2 steps. We
remark that, however, there is no easy way to figure out the
state-dependent sequence of steps for an ensemble of states with
specific initial fidelity $F$. The result is also compatible with
the theoretical limit of $F>0.3$ which is given by the PPT
criterion \cite{PPT1,PPT2} for the Werner-like states
\cite{dctprl99,dcpra00}.

Let us now explain Murao et al's protocol \cite{mppvk98} in the
modern stabilizer formulation: In fact, their protocol is to
apply post-selections based on bit-flip error syndromes and
phase-flip error syndromes separately. The P2 step corresponds to
a bit-flip error syndrome measurement followed by post-selection,
while the P1 step corresponds to phase-flip error syndrome
measurement followed by post-selection. Here the stabilizer for
the GHZ state is CSS-like and errors can be corrected by two
steps together. We call a distillation procedure to be CSS-based
distillation if all the measurement operators used are of either
$X$-type or $Z$-type. Murao et al's protocol is thus a CSS-based
GHZ state distillation.

\vspace{0.3cm} \noindent \textbf{Remark 5:} Recurrence protocols
have asymptotically zero yields when the required fidelity go to
one. This is because after P1 or P2 step, the residual states may
have a higher fidelity but losing at least one half of the
states. Thus the yield for the recurrence protocol is zero in the
asymptotic sense. Therefore, at a high noise level, one should
start with a number of rounds of the recurrence protocol and
switch to a hashing protocol when the hashing protocol starts to
work. For this reason, it is important to study both hashing and
recurrence protocols.

\section{CSS states}
\noindent
In this section, we generalize our results on the GHZ state to a
general CSS state. We first define a CSS state and show that,
just like the GHZ state, a CSS state can be distilled by the
various aforementioned protocols (recurrence and hashing). We
derive the yield for the hashing protocol. Finally, we show that
a CSS state is equivalent to a bipartite (i.e., two-colorable)
graph state, a subject of recent attention.

\subsection{Distillation of CSS-states}
\noindent
\label{ss:CSSstates} A CSS-state is basically a CSS-code
\cite{CSS1,CSS2} where the number of encoded qubits is zero. For
instance, an encoded $\ket{0}$ state of a CSS code is a
CSS-state. More formally, we have the following definition.

{\it Definition:} {\bf CSS states.} A CSS-state is a $+1$
eigenstate of a complete set of (commuting) stabilizer generators
such that each stabilizer element is of X-type or Z-type only.

{\it Example:} A GHZ-state is a CSS-state with stabilizer
generators $XXX$, $ZZI$ and $ZIZ$.

{\bf Claim~1}: Consider the distillation of a multi-partite
CSS-state, given $N$ noisy versions of such a state. Consider a
complete set of CSS generators (i.e., each of X-type or Z-type).
Suppose we label its simultaneous eigenstate by its simultaneous
eigenvalues $\ket{\hat{b} , \hat{p}}$ where $\hat{b}= \{b_1, b_2,
\cdots, b_m\}$ is a vector that denotes the tuples of Z-type
eigenvalues and $\hat{p}= \{p_1, p_2, \cdots, p_n\}$ is a vector
that denotes the tuples of X-type eigenvalues. Consider a pair of
multi-partite states. We claim that, under multilateral quantum
XOR gates (MXOR), the state of the pair evolves as follows
\begin{equation}
\text{MXOR} \left[ \ket{\hat{b^1} , \hat{p^1}} \ket{\hat{b^2} ,
\hat{p^2}} \right] = \ket{\hat{b^1} , \hat{p^1}+\hat{p^2}
}\ket{\hat{b^1}+ \hat{b^2} , \hat{p^2}}.
\end{equation}
That is to say, that the bit-flip errors propagate forward and
the phase errors propagate backwards.

Claim~1 follows from a standard result in quantum error
correction. It is easy to understand from the evolution of Pauli
operators acting on individual qubits. We will skip its proof
here.

The upshot of Claim~1 is that much of what we learn from GHZ can
be directly generalized to a general multi-partite CSS-state. For
instance, we can apply the hashing protocols and sub-routines P1
and P2 to a general CSS state. In what follows, we will discuss
this point in more detail.

By using random hashing (in Z and followed by in X), one can
easily show that the yield is
\begin{equation}
D_h > 1 - \max_{i} \left[H(b_i)\right] - \max_{j}
\left[H(p_j)\right], \label{eqn:naivehashingCSS}
\end{equation}
where the maximum over $i$ is over all bit-flip eigenvalues,
$b_i$, and the maximum over $j$ is over all phase eigenvalues,
$p_j$.

As in the case of GHZ, we can apply the trick in
\cite{losixstate01} to improve the result to
\begin{equation}
D_h > 1 - \max_{i} \left[H(b_i)\right] - \max_{j}
\left[H(p_j|\hat{b})\right].
\end{equation}
This is so because the mutual information $I (\hat{b}, \hat{p})$
between the bit-flip and phase syndromes is non-zero. Therefore,
learning the bit-flip error pattern allows one to reduce the
entropy for the phase error pattern to the {\it conditional}
entropy given the bit-flip error pattern.

Alternatively, one can hash in X first and then hash in Z.
Therefore, we also have
\begin{equation}
D_h > 1 - \max_{j} \left[H(p_j)\right] - \max_{i}
\left[H(b_i|\hat{p})\right].
\end{equation}
Combining the two results, we have
\begin{eqnarray}
D_h > \max \Big( 1 - \max_{i} \left[H(b_i)\right] -
\max_{j} \left[H(p_j|\hat{b})\right] ,
1 - \max_{j} \left[H(p_j)\right] - \max_{i}
\left[H(b_i|\hat{p})\right] \Big) .\label{eqn:hashingCSS}
\end{eqnarray}
As will be shown below, CSS-states are equivalent to bipartite
states. Our hashing result (\ref{eqn:hashingCSS}) is an
improvement over the prior art result in
\cite{smolincm2002,graph}.

In principle, for some states, we will be able to improve the
result further. As an example, suppose $\hat{b}$ consists of only
$b_1 $ and $b_2$ and $H(b_2) > H(b_1)$ and $I(b_2; b_1) > 0$.
Then, we can improve the yield to
\begin{eqnarray}
 D_h'& > & \max \Big(1 - \max \left\{H(b_1), H(b_2| b_1) \right\} -
\max_{j} \left[H(p_j|\hat{b})\right], \nonumber \\
&& \;\;\;\;\;\;\;\; 1 - \max_{j} \left[H(p_j)\right] - \max_{i}
\left[H(b_i|\hat{p})\right] \Big). \label{eqn:newhashingCSS}
\end{eqnarray}
Here, by a similar argument to what we used before for a GHZ
state, we make use of the non-trivial mutual information $I(b_2;
b_1)$ between $b_1$ and $b_2$ to reduce the rounds of hashing
needed.

We note that one can apply the subroutines P1 and P2 to any
CSS-state. As in the case of GHZ, this will allow the successful
distillation of CSS-state at higher error rates than what is
possible with only a hashing protocol.

\section{Equivalence of CSS-states and bipartite graph states}
\noindent
Distillation of another type of states, so-called graph states,
have also been discussed in the literature. In particular,
D\"{u}r, Aschauer, and Briegel \cite{graph} have recently found
systematic procedures for distilling a special class of graph
states---those that are bipartite (i.e., 2-colorable). In what
follows, we will describe bipartite graph states and show that
they are essentially equivalent to CSS-states. An undirected
graph is specified by a set of vertices, $V$ and edges, $E$. A
graph is said to be bipartite (i.e., 2-colorable) if the set of
vertices can be decomposed into two sets, say $L$ (left) and $R$
(right) such that an edge is only allowed to connect a $L$ vertex
with a $R$ vertex (but not between two $L$ (or $R$) vertices).

A graph state is a multi-partite state where each vertex, $v_j$,
represents a qubit. A graph state has stabilizer generators of
the form $K_j = X_j \Pi_{(j,k) \in E} Z_k$.

One part (Part B) of the following Claim is due to Eric Rains.

{\bf Claim~2}: Bipartite (i.e, two-colorable) graph states are
equivalent to CSS-states. (Since the concepts of
bipartite/two-colorable and CSS are {\it not} invariant under
local unitary transformations, one has to be careful in stating
the claim. More precisely, Claim~2 says that any state that can
be represented by a bipartite/two-colorable graph state can be
written as a CSS-state and vice versa.)

{\it Proof}: (Part~A: Bipartite implies CSS.) Given a bipartite
graph state. Let us consider the following operation. We apply a
Hadamard transformation to all ``left'' vertices  and the
identity operator to all ``right'' vertices. Now, for any left
vertex, $v_j$, the associated stabilizer generator becomes
$Z$-type (because $X_j \to Z_j$ and any vertex $v_k$ connected to
$v_j$ is a right vertex and so $Z_k \to Z_k$). On the other hand,
for any right $v_j$, the associated stabilizer generator becomes
$X$-type (because now $X_j \to X_j$ itself and $Z_k \to X_k$, as
$v_k$, being connected to $v_j$, must be a left vertex).

(Part~B: CSS implies bipartite. The result is due to Eric Rains.)
Given a CSS-state. Consider the associated classical code, $C$.
Its parity check matrix in canonical form is $H_C = \left[ I|A
\right]$ where $I$ is say a $k$ by $k$ matrix and $A$, a $k$ by
$n-k$ by matrix. Now, the parity check matrix for its dual code,
$C^{\perp}$, is given by $H_{C^{\perp}}= [A^T |I ]$ where $A^T$
is an $n-k$ by $k$ matrix and $I$ an $n-k$ by $n-k$ matrix. Note
that a CSS-state has a complete set of stabilizer generators.
This implies that the parity check matrix of a CSS-state must be
of the form
\begin{eqnarray}
H &=&\left(
\begin{array}{cc}
H_C & 0 \\
0   & H_C^{\perp}
\end{array}
\right)
\nonumber\\
&=&\left(
\begin{array}{cccc}
I & A & 0 &  0 \\
0 & 0 & A^T & I
\end{array}
\right) .
\nonumber\\
\end{eqnarray}
We now apply Hadamard transforms on the last $n-k$ qubits. This
will inter-change between $X$ and $Z$, thus mapping the parity
check matrix to be
\begin{equation}
H =\left(
\begin{array}{cccc}
I & 0 & 0 &  A \\
0 & I & A^T & 0
\end{array}
\right) .
\end{equation}

Notice that the above parity check matrix indeed represents a
bi-partite graph. Let us regard the first $k$ qubits as left
vertices and the remaining $n-k$ qubits as right vertices. Note
that there is indeed no edge between vertices on the same side.
And, $A$ represents the adjacent matrix of the graph.~Q.E.D.

Claim~2 establishes the equivalence of two different mathematical formulations:
CSS-states and bipartite graph states. It means that much of what we have
learnt about the distillation of bipartite/two-colorable graph states through
the work of D\"{u}r, Aschauer, and Briegel \cite{graph} can be interpreted in
the more systematic formulation of CSS-states. In particular, it is natural to
consider the bit-flip and phase error patterns separately and consider their
propagations in quantum computational circuit. From Claim~2, we learn that
Claim~1, which originally refers to CSS states, can be applied directly to any
bipartite graph states. Moreover, the improvement that we have found in
(\ref{eqn:newhashingCSS}) over (\ref{eqn:naivehashingCSS}), in fact, applies to
bipartite graph states. Notice that the basic primitives in our protocol are
the same as in \cite{graph}. However, the yield in \cite{graph} will go to zero
when the required fidelity go to one seen from Remark 5, which is also pointed
out in a recent progress \cite{graph0}. In \cite{graph0}, a one-way hashing
protocol is further proposed, that is essentially a generalization of the one
developed in \cite{smolincm2002} and thus is identical to
Eq.~(\ref{eqn:naivehashingCSS}). Therefore, to achieve a given fidelity, the
numbers of rounds that one needs will be smaller using our method than the one
in \cite{graph,graph0}, by combining the recurrence protocol and our improved
hashing protocol that gives yield of Eq.~(\ref{eqn:newhashingCSS}).

\subsection{Distillation of non-CSS states}
\noindent
So far, our discussion has focussed on CSS-states (i.e.,
2-colorable graph states). It will, thus, be interesting to
investigate the distillation of non-CSS-states. We believe that
this is a highly non-trivial question for the following reasons.

First of all, as noted earlier, whether a state is in a CSS form
or not is not invariant under local unitary transformations. For
this reason, given a general stabilizer state, there is no known
efficient algorithm to determine whether it can be realized as a
CSS-state. Second, when a state is known to be non-CSS, it is
highly unclear what the basic primitives for distillation should
be.

However, one possible approach---cut and re-connect---has already
been discussed in \cite{graph,graph1}. Their idea is by
performing some local measurements, the parties can {\it cut} a
graph state into a sub-graph state. Suppose one is given 2N
copies of some noisy version of a non-bipartite graph states. One
might imagine cutting $N$ copies in a specific way to generate
some sub-graph that is bipartite (and, therefore, CSS-like). They
can then distill this CSS sub-graph state. They also cut the
other $N$ copies in another way to generate another sub-graph
that is bipartite. They can then distill this other CSS sub-graph
state. Afterwards, they can re-connect these two subgraphs by
performing some (say Bell-type) measurements. More generally,
they can apply this cut-and-reconnect strategy iteratively. The
extreme case of this strategy is to cut everything into EPR-pair
states. One then distills only bi-partite EPR pair states and
re-connect them by teleportation. In our opinion, it will be
interesting to study this cut and re-connect strategy in more
detail.

Recently, the distillation of W-state (a non-CSS state) has been
discussed in \cite{Wstatedistill}

\section{Multi-party Quantum Cryptography}
\noindent
\subsection{Adversarial Model}
\noindent
The key simplifying assumption we make here is that, at least in
the first phase of the protocol, the three participating parties,
Alice, Bob, and Charlie, will always perform their tasks
honestly. The case where some of the participants may be
dishonest even in the first phase of the protocol is an
interesting subject that deserves future investigations.

What threats are we trying to address here? We consider an
eavesdropper, Eve, who may actively tamper with quantum channels
and also passively monitor all classical communications. Our goal
is to prevent Eve from learning secrets in the protocol. In the
second phase of the protocol, some of the participants are
allowed to be dishonest.

In summary, Eve is active, and is collaborating with Bob (or
Charlie), who only cheats ``passively'' (i.e., honest but
curious).

So, what multi-party cryptographic tasks do we consider? We will
consider three tasks, namely (a) conference key agreement, (b)
quantum sharing of classical secrets and (c) third-man
cryptography.

\subsection{Conference Key Agreement}
\noindent
A well-known problem in classical cryptography is secret
broadcasting and conference key agreement \cite{schneier}.
Suppose Alice would like to broadcast a message securely to only
Bob and Charlie, in the presence of an eavesdropper, Eve. One
simple method to achieve it is for Alice, Bob and Charlie to
first share a common conference key, $k$. Then, Alice can use the
key, $k$, as a one-time-pad for encrypting her message.

The goal of conference key agreement is precisely for three
parties, Alice, Bob, and Charlie, to obtain a common random
string of numbers, known as the conference key, $k$, and to
ensure that $k$ is secure from any eavesdropper, Eve. Consider
the following scenario: Alice, Bob and Charlie initially share
some quantum states that either are prepared by Eve or are
prepared by Alice, but have to go through Eve's channels before
reaching Bob and Charlie. We will specialize to
``prepare-and-measure'' type of protocols, mentioned in our
introduction.

To secure their classical communications, we assume that each
pair of the three parties, Alice, Bob and Charlie, also share a
pairwise authenticated classical channel. This can be achieved by
requiring that they share a small amount (logarithmic in the
amount of classical communications) of authentication key. For
simplicity, we will assume perfect authentication. Besides
tampering with quantum channels, Eve can listen to all classical
communications passively.

Since Alice, Bob and Charlie are in the same boat, we think it is
fair to assume that they will always execute the protocol
honestly.

\subsection{Quantum Sharing of Classical Secrets}
\noindent
The second protocol is the quantum sharing of classical secrets
\cite{hbb99}, which has been demonstrated experimentally in a
recent experiment \cite{expQSCS}. Whereas Ref.~\cite{hbb99} deals
with perfect GHZ states, we consider noisy GHZ states. Suppose a
President, Alice (A), has the password for a bank vault and he
would like to divide up this secret password between two Vice
Presidents, Bob (B) and Charles (C), in such a way that neither B
nor C alone knows anything about the password and yet when B and
C come together, they can re-generate the password. Suppose
further that A, B and C live very far away from each other. An
eavesdropper, Eve may try to intercept the communications during
the secret sharing phase.

Here, we make a rather strong assumption that in the distribution
and verification phase of the quantum signals, the three
parties---A, B, and C---are trustworthy. However, an adversary,
Eve, can tamper with quantum channels. Besides, Eve may eavesdrop
passively on all classical communications during the verification
step.

{\it Phase One} (Distribution and verification phase): In the
first phase of the protocol, the three parties (trusted for the
moment) use quantum states to perform the sharing of binary
correlations, $X_A$, $X_B$ and $X_C$ satisfying
\begin{equation}
X_A \oplus X_B \oplus X_C= 0 ,
\end{equation}
between them. Alice, Bob and Charlie perform local, individual
measurements on qubits and also classical computations and
classical communications between each other. We assume that all
classical communications are authenticated by standard
unconditionally secure method of authentication. Finally, either
Alice, Bob and Charlie abort the protocol or they have the
confidence that they have shared those correlations.

{\it Phase Two} (Broadcast Phase): Suppose Alice would like to
share a secret bit $a$ with Bob and Charles. We assume Alice has
a broadcast channel. Alice broadcasts $b= a \oplus X_A$. Note
that if Bob and Charles get together, they can compute $X_B
\oplus X_C$ and thus, $b \oplus X_B \oplus X_C = a \oplus X_A
\oplus X_B \oplus X_C = a$. Therefore, Bob and Charles can
recover the secret $a$, if they come together. However, neither
Bob or Charles on his own has any information at all about the
bit, $a$. In summary, Alice, Bob and Charlie, have achieved
sharing of classical secrets.

In summary, the goal of quantum sharing of classical secrets is
to ensure that an eavesdropper Eve cannot learn about the shared
secrets. We assume that Alice, Bob and Charles, are trustworthy
in Phase One of the protocol.

\subsection{Third-Man Cryptography}
\noindent
Our quantum secret sharing protocol can be modified into a
``third-man cryptography'' protocol. This will be discussed in
Subsection \ref{ss:thirdman}

\section{3-party conference key agreement with noisy GHZ states}
\noindent
In this section, we apply the ideas of multi-partite entanglement
distillation to a conference key agreement scheme. We consider a
prepare-and-measure protocol where a preparer distributes $n$
trios of qubits to the three parties.

Note that, if one of the parties, say Alice, is actually the
preparer of the multi-partite state, in a prepare-and-measure
protocol, she is allowed to pre-measure her subsystem. By doing
so, she has projected an $N$-partite entangled state into one of
the various $(N-1)$-partite entangled state. Therefore, the
protocol can be implemented with only $(N-1)$-party entanglement.
The above conversion idea is similar to the idea in
\cite{Shor-Preskill} and \cite{GLIEEE03}.

How do we prove the security of our prepare-and-measure protocol?
We start with a CSS-based GHZ state distillation protocol and try
to convert it to a ``prepare and measure" protocol. We draw our
inspiration from \cite{Shor-Preskill,GLIEEE03} where security of
BB84 is proven by conversion from an entanglement distillation
protocol (EDP). Specifically, Ref.~\cite{GLIEEE03} shows that for
an EDP to be converted to a prepare-and-measure protocol (namely,
BB84), it can include bit-flip error syndrome measurement and
post-selection as well as phase error syndrome measurement.
However, following the discussion in \cite{GLIEEE03}, the
post-selection of surviving systems based on phase error
syndromes is strictly forbidden, if an EDP is convertible to
BB84. This is because, without a full-blown quantum computer, it
is impossible for a participant to hold measurement outcomes
simultaneously in two different bases (bit-flip and phase) for
individual qubits. Without those measurement outcomes, a
participant cannot know what {\it conditional} or {\it
post-selected} actions to take for bit-flip and phase error
detections simultaneously.

Let us first look at the original protocol by Murao et al
\cite{mppvk98}. The P2 step can be considered as a bit-flip error
syndrome measurement and post-selection step and as such it is
allowed in a prepare-and-measure protocol. Then, the P1 step,
which involves post-selection based on phase error syndromes, is
strictly forbidden in a ``prepare-and-measure" protocol
\cite{GLIEEE03}. Therefore, Murao et al's protocol cannot be
converted to a ``prepare and measure" protocol. This is why we
have to search for a new primitive.

\begin{figure} [htbp]
%\vspace*{13pt}
\centerline{\epsfig{file=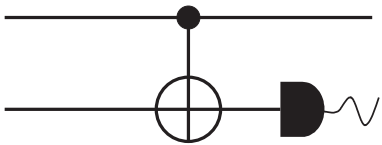, width=5cm}} %100 percent
\vspace*{13pt}
\caption{\label{fig4}B step (bit-flip error detection) procedure for
conference key agreement protocol.}
\end{figure}

\begin{figure} [htbp]
%\vspace*{13pt}
\centerline{\epsfig{file=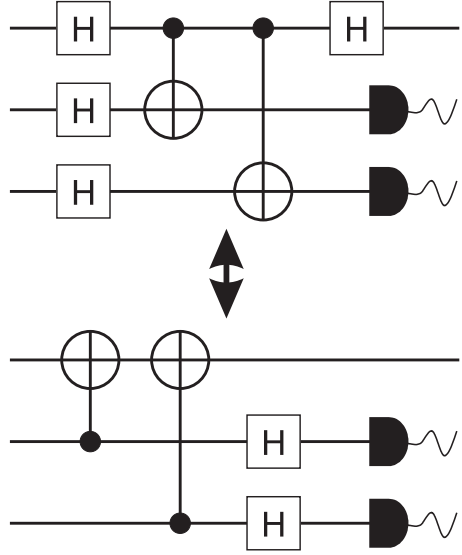, width=6cm}} %100 percent
\vspace*{13pt}
\caption{\label{fig5}P step (phase-flip error correction) procedure for
conference key agreement protocol. In a prepare-and-measure
protocol for conference key agreement, each of Alice, Bob and
Charlie simply takes the parity $(Z_{A}+Z_{B}+Z_{C})\mod 2$ of
their own three particles, which can be done locally and no
classical communication is needed.}
\end{figure}

Let us construct some subprotocols. For bit-flip error detection
(B step), we use the same procedure as the P2 step of Murao et al
\cite{mppvk98} as shown in Fig.~\ref{fig4}. As for phase error
correction (P step), we design a procedure using the idea of
Gottesman and Lo \cite{GLIEEE03} as shown in Fig.~\ref{fig5}, but
apply it to the multi-partite case. \vspace{0.5cm}

\textbf{Our multi-partite B step:} Using the formulation of
stabilizer, this step can be
reformulated by the following transformation of the two GHZ-like states
\begin{eqnarray}
[(p,i_{1},i_{2}),(q,j_{1},j_{2})]
\overset{\text{applying BXOR}}{\longrightarrow }(p\oplus
q,i_{1},i_{2}),(q,i_{1}\oplus j_{1},i_{2}\oplus j_{2}),
\end{eqnarray}
where $(p,i_{1},i_{2}),(q,j_{1},j_{2})$ denote the phase bit and
amplitude bits for the first and second GHZ-like states. If
$i_{1}\oplus j_{1}=i_{2}\oplus j_{2}=0 \mod 2$, we keep the first
GHZ state, otherwise discard all the two states. In a
prepare-and-measure conference key agreement protocol,
this corresponds to the prescription that we keep the first trio iff
$M_{A}=M_{B}$ and $M_{A}=M_{C}$ (i.e., Alice, Bob and Charlie get
the same
measurement outcome). This step changes the 8 elements of diagonal entries
$(p_{000},p_{100},p_{011},p_{111},p_{010},p_{110},p_{001},p_{101})^{t}$
to
\begin{equation}
\left(
\begin{array}{c}
p_{000}^{2}+p_{100}^{2} \\
2p_{000}p_{100} \\
p_{011}^{2}+p_{111}^{2} \\
2p_{011}p_{111} \\
p_{010}^{2}+p_{110}^{2} \\
2p_{010}p_{110} \\
p_{001}^{2}+p_{101}^{2} \\
2p_{001}p_{101}
\end{array}
\right) /P_{pass},
\end{equation}
where $P_{pass}=\big(%
(p_{000}+p_{100})^{2}+(p_{001}+p_{101})^{2}+(p_{010}+p_{110})^{2}+(p_{011}+p_{111})^{2}%
\big)$ is the probability for the survived subset.

\textbf{Our multi-partite P step:} Using the formulation of
stabilizer, this step (shown in Fig.~\ref{fig5}) can be
reformulated by the following transformation of the three
GHZ-like states
\begin{align}
&[(p,i_{1},i_{2}),(q,j_{1},j_{2}),(r,k_{1},k_{2})] \nonumber \\
&\overset{\text{applying 1st BXOR}}{\longrightarrow
} [(p,i_{1}\oplus j_{1},i_{2}\oplus j_{2}),
(p\oplus q,j_{1},j_{2}),(r,k_{1},k_{2})] \nonumber \\
&\overset{\text{applying 2nd BXOR}}{\longrightarrow }
[(p,i_{1}\oplus j_{1}\oplus k_{1},i_{2}\oplus j_{2}\oplus
k_{2}),(p\oplus q,j_{1},j_{2}),(p\oplus r,k_{1},k_{2})].
\label{pconf}
\end{align}
If $p\oplus q=p\oplus r=1\mod 2$, we apply $p\longrightarrow p\oplus 1%
\mod 2$, otherwise keep the first GHZ-like state invariant. Note
that P can also be performed locally by each party, which regards
the circuit as implementing a 3-qubit phase error correction
code. The important point to note is that no post-selection on
surviving systems is needed in our multi-partite P step.

Here the condition exactly corresponds to the fact that, if an
odd number of the source bits are measured to be in the same
state $|1\rangle $ both in the second and in the third GHZ-like
states, we just apply a $Z$ operation in Alice's qubit for the
first GHZ-like states. Otherwise, just keep the first GHZ-like
state invariant. The same as for the B step, to obtain the
density matrix after the P step, one just needs to count and sum
all the probability for the residual state according to
Eq.~(\ref{pconf}). For example, if the three initial quantum states
in Fig.~\ref{fig5} are in
$(p,i_{1},i_{2}),(q,j_{1},j_{2}),(r,k_{1},k_{2})$ with
probabilities of
$p_{p,i_{1},i_{2}},p_{q,j_{1},j_{2}},p_{r,k_{1},k_{2}}$,
respectively, then the final state will be $(p,i_{1}\oplus
j_{1}\oplus k_{1},i_{2}\oplus j_{2}\oplus k_{2})$ with
probability $p_{p,i_{1},i_{2}} p_{q,j_{1},j_{2}}
p_{r,k_{1},k_{2}}$. Sum all the possibilities for such
combinations (in our case, one has $8\times 8\times 8=512$
different probabilities combinations for three initial quantum
states), one arrives a final GHZ-diagonal state, which we skip
the verbose expression here.

\subsection{Conversion to a prepare-and-measure protocol}
\noindent
\label{ss:reduction1} We now argue that the aforementioned
entanglement distillation subprotocol (P step) can be converted
to a prepare-and-measure protocol for conference key agreement.
This is in the spirit of \cite{GLIEEE03}. Recall that in a
prepare-and-measure protocol, the three parties, Alice, Bob and
Charlie, are presented with an ensemble of noisy GHZ states. They
perform local measurements on individual qubits and then
classically compute and classically communicate with each other.
Their goal is to share a common string of random number between
the three parties so that an eavesdropper will have an negligible
amount of information on it. Such a common secret random number
can then be used as a key for a conference call between the three
parties.

In the entanglement distillation picture, for conference key
agreement, Alice, Bob and Charlie do not need to perform phase
error correction. They only need to prove that, phase error
correction would have been successful, if they had performed it.
In a prepare-and-measure protocol for conference key agreement,
by the second part of Fig.~\ref{fig5}, each of Alice, Bob and
Charlie simply takes the parity $(Z_1+Z_2+Z_3)\mod 2$ of their
own three particles. In other words, each computes locally the
parity of his/her three measurement outcomes. No classical
communication is needed. This has been discussed in
\cite{GLIEEE03}.

Moreover, following \cite{GLIEEE03}, we can apply the same
conversion idea to any concatenated protocols involving B steps,
P steps and following by a hashing protocol. This conversion
result means that we can obtain secure protocols for conference
key agreement by considering the convergence of GHZ distillation
protocols involving those operations.

In what follows, we consider only Werner-like states. By direct
numerical calculation, we can verify that our scheme can distill
GHZ states with nonzero yield whenever $F\geq 0.3976$ by some
state-dependent sequence of B and P steps, and then change to our
random hashing method if it works. We find that a sequence of B
and P steps BBBBB for $F=0.3976$, which is optimal for any
sequence with at most 5 steps (consisting of only B or P) plus
immediate random hashing. Compared with the random hashing method
of Maneva and Smolin which works only when $F\geq 0.8075$ and our
improved random hashing method which works only when $F\geq
0.7554$, our new protocol gives dramatic improvement by using
2-way classical communications. We just execute some
state-dependent sequence of B and P steps, and then change to the
hashing method once the effective error rate has decreased to the
point where our improved hashing method actually works. These
results imply that Alice, Bob and Charlie can achieve secure
conference key agreement if they share a generalized Werner state
with a fidelity $F\geq 0.3976$. It should be remarked that the
sequence BBBBB works only for Werner state with initial fidelity
around 0.3976. A different initial fidelity requires a different
optimal sequence plus immediately random hashing. Also a
different sequence length causes different optimal sequence.
Similar to the case of distilling directly GHZ state using P1 and
P2 steps shown in \cite{dcpra00}, there is no easy way to derive
an optimal sequence as well for our protocols. However, we remark
that in practice 5 steps of P and B combinations is reasonable
sequence length and is easy to be optimized with a standard
computer by a simple program. The three parties just need to
estimate reliably the shared GHZ-basis diagonal states (after
depolarization procedure in Section II.C) and then obtain an
optimal sequence for them by performing the computer search. Then
the three parties execute B step by CCCC.  Whenever there is a P
step, they take the parity $(Z_1+Z_2+Z_3)\mod 2$ of their own
three particles locally. No classical communication is needed for
a P step. Even when the shared GHZ-basis diagonal state is not a
Werner state, the same analysis can be applied to obtain some
optimal sequence of B and P steps combining with our hashing
method and see if one can succeed to make conference key
agreement.

\subsection{Experimental Implementations}
\noindent
\label{ss:experiments} For practical implementations, we remark
that, if one of the parties, say Alice, is the preparer of the
state, then in a prepare-and-measure protocol she has the option
to pre-measure her subsystem. By doing so, she has converted a
protocol that involves $N$-party entanglement to one that
involves only $(N-1)$-party entanglement.

Consider the case when $N=3$. The above discussion means that
three-party conference key agreement can be implemented with only
bi-partite entangled states. More concretely, imagine that Alice
prepares a perfect GHZ state and measures her qubit along the
$Z$-axis. After her measurement, Bob and Charlie's state is in
either $\ket{00}$ or $\ket{11}$ (with equal probabilities).
Similarly, suppose Alice measures her qubit along the $X$-axis.
After her measurement, Bob and Charlie's state is in either $
{1/\sqrt{2}} ( \ket{00} + \ket{11} )$ or $ {1/\sqrt{2}} (
\ket{00} - \ket{11} )$ (with equal probabilities). Similarly,
suppose Alice measures her qubit along the $Y$-axis. After her
measurement, Bob and Charlie's state is in either $ {1/\sqrt{2}}
( \ket{00} +i \ket{11} )$ or $ {1/\sqrt{2}} ( \ket{00} -
i\ket{11} )$ (with equal probabilities).

In summary, Alice could implement conference key agreement by
simply preparing one of the six states $\ket{00}$, $\ket{11}$, $
{1/\sqrt{2}} ( \ket{00} + \ket{11} )$, $ {1/\sqrt{2}} ( \ket{00}
- \ket{11} )$, $ {1/\sqrt{2}} ( \ket{00} +i \ket{11} )$, or $
{1/\sqrt{2}} ( \ket{00} - i\ket{11} )$ and sending the two qubits
to Bob and Charlie respectively through some quantum channels.
Notice that the six states listed above only involve bi-partite
entanglement and as such can be readily prepared by parametric
down conversion sources. For this reason, it is feasible to
demonstrate experimentally our conference key agreement scheme.

While we have focussed our discussion on the three-party case, we
remark that the basic concepts apply directly to the $N >3$-party
case. Moreover, a prepare-and-measure protocol involving
$N$-parties can be implemented with $(N-1)$-partite entanglement.
We shall skip the details here.

\section{Secret sharing in a noisy channel}
\noindent
\subsection{Quantum Secret Sharing}
\noindent
In a classical secret sharing scheme, some sensitive data is
divided up into shares among a number of people so that it can be
re-constructed if and only if a sufficiently large number of
people get together. For instance, in a classical
$(k,n)$-threshold scheme, the secret is divided among $n$ parties
such that when $k$ or more parties get together, they can
re-construct the secret. On the other hand, if less than $k-1$
parties get together, then they have absolutely no information on
the secret.

The idea of $(k,n)$-threshold quantum secret sharing scheme was
first proposed in \cite{CGL99}: A preparer presents the dealer
with an unknown quantum state. The dealer then divides up the
unknown state among $n$ parties. The quantum no-cloning theorem
demands that $2k > n$ for the existence of a $(k,n)$-threshold
quantum secret sharing scheme. This is the only requirement for
the existence of a quantum secret sharing scheme.

It was then shown in \cite{gottesman00} that for any access
structure that does not violate the quantum no-cloning theorem, a
quantum secret sharing scheme exists. We remark the construction
in \cite{CGL99,gottesman00} makes use of CSS codes.

\subsection{Quantum sharing of classical secrets}
\noindent
Another related concept---quantum sharing of classical
secrets---has also been introduced \cite{hbb99}. Here, the goal
is to use quantum states to share a {\it classical} secret
between multiple parties and to ensure that no eavesdropper can
learn useful information by passive eavesdropping. It has been
shown \cite{hbb99} that in the absence of noises, one can
construct a secure secret sharing scheme using perfect GHZ state
by just measuring \emph{along $X$ basis} between Alice, and
cooperating Bob and Charlie. The key point is that the GHZ state
is a $+1$ eigenstate of $XXX$. So, by measuring the $X$-basis,
Alice, Bob and Charlie's measurement outcomes will satisfy a
classical constraint of $X_A + X_B + X_C = 0 \mod 2$. Note that
individually, each of Bob and Charlie has no information on
$X_A$. But, by getting together, Bob and Charlie can obtain $X_B
+ X_C \mod 2$ and, therefore, obtain $X_A$. Suppose Alice now
would like to share a bit value, $b$, with Bob and Charlie. She
can broadcast $X_A + b$.

Now, let us consider the case of noisy GHZ states. We want to
develop a GHZ distillation protocol which can be converted to a
``prepare and measure" secret sharing scheme. Such a scheme is
technologically less challenging than a full-blown quantum
computing scheme and yet one can analyze its security by using
our ideas for GHZ distillation protocols presented in this paper.

\begin{figure} [htbp]
%\vspace*{13pt}
\centerline{\epsfig{file=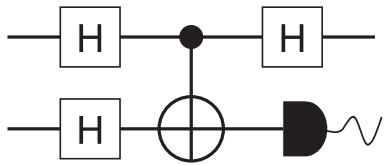, width=5cm}} %100 percent
\vspace*{13pt}
\caption{\label{fig6}B' step (bit-flip error detection) procedure for secret
sharing.}
\end{figure}

\begin{figure} [htbp]
%\vspace*{13pt}
\centerline{\epsfig{file=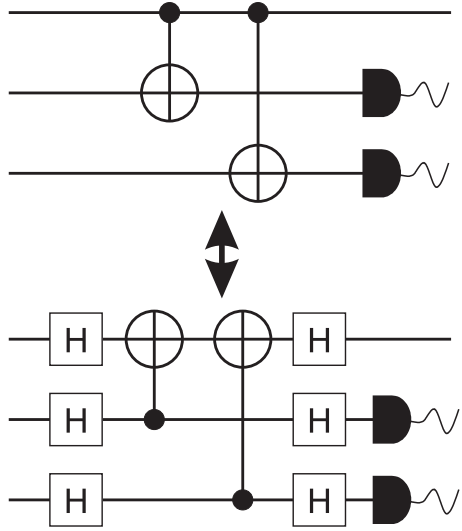, width=6cm}} %100 percent
\vspace*{13pt}
\caption{\label{fig7}P' step (phase-flip error correction) procedure for
secret sharing. In a prepare-and-measure protocol for secret
sharing, each of Alice, Bob and Charlie simply takes the parity
$(X_{A}+X_{B}+X_{C})\mod 2$ of their own three particles, which
can be done locally and no classical communication is needed.}
\end{figure}

Note that, since the final measurement is now done along the
X-axis, the bit-flip measurements now correspond to measurements
along X. The readers should bear this point in mind to avoid any
future confusion. Therefore, for the bit-flip error detection
code (B' step), we use the same procedure as the P1 step of Murao
et al \cite{mppvk98} as shown in Fig.~\ref{fig6}. As for the
phase error correction (P' step), we design a procedure similar
to the idea of Gottesman and Lo \cite{GLIEEE03} as shown in
Fig.~\ref{fig7}. \vspace{0.5cm}

\textbf{Our P' step for secret sharing:} Using the formulation of
stabilizer, this step (shown in Fig.~\ref{fig7}) can be
reformulated by the following transformation
of the three GHZ-like states
\begin{align}
&[(p,i_{1},i_{2}),(q,j_{1},j_{2}),(r,k_{1},k_{2})] \nonumber \\
&\overset{\text{applying 1st BXOR}}{\longrightarrow } [(p\oplus
q,i_{1},i_{2}),(q,i_{1}\oplus j_{1},i_{2}\oplus j_{2}),(r,k_{1},k_{2})] \nonumber \\
&\overset{\text{applying 2nd BXOR}}{\longrightarrow }
[(p\oplus q\oplus r,i_{1},i_{2}),(q,i_{1}\oplus
j_{1},i_{2}\oplus j_{2}),(r,i_{1}\oplus k_{1},i_{2}\oplus k_{2})].
\label{pss}
\end{align}
If $i_{1}\oplus j_{1}=i_{1}\oplus k_{1}=1 \mod 2$, we apply
$i_{1}\longrightarrow i_{1}\oplus 1 \mod 2$. If $i_{2}\oplus
j_{2}=i_{2}\oplus k_{2}=1 \mod 2$, we apply $i_{2}\longrightarrow
i_{2}\oplus 1 \mod 2$. Otherwise keep the first GHZ-like state
invariant. Note that the phase error correction procedure can be
performed locally by each party because the circuit diagram in
Fig.~\ref{fig7} is simply a simple three-qubit majority vote
code. The condition exactly corresponds to the fact that, if
Alice's and Bob's measurement outcomes disagree for both the
second and the third GHZ-like states, we just apply a $X$
operation in Bob's qubit for the first GHZ-like states, and that,
if Alice's and Charlie's measurement outcomes disagree for both
the second and the third GHZ states, we just apply a $X$
operation in Charlie's qubit for the first GHZ-like states.
Otherwise, just keep the first GHZ-like state invariant. Follow
the similar counting procedure as done for the P step of
conference key agreement, one can knows the final states exactly
by straightforward calculation according to Eq.~(\ref{pss}).

We will now argue that this phase correction protocol can be
converted to a ``prepare-and-measure'' protocol for quantum
secret sharing. In a prepare-and-measure protocol for secret
sharing, each party simply computes the parity of his/her own
three measurement outcomes locally. No classical communication is
needed.

\textbf{Our B' step for secret sharing:} Again using the
formulation of stabilizer, this step (shown in Fig.~\ref{fig6})
can be reformulated by the following
transformation of the two GHZ-like states
\begin{equation}
[(p,i_{1},i_{2}),(q,j_{1},j_{2})]
\overset{\text{applying BXOR}}{\longrightarrow }(p,i_{1}\oplus
j_{1},i_{2}\oplus j_{2}),(p\oplus q,j_{1},j_{2}).
\end{equation}
If $p\oplus q=0 \mod 2$, we keep the first GHZ-like state,
otherwise discard all
the two states. This step changes the 8 elements of diagonal entries
$(p_{000},p_{100},p_{011},p_{111},p_{010},p_{110},p_{001},p_{101})^{t}$ to
\begin{equation}
\left(
\begin{array}{c}
p_{000}^{2}+p_{001}^{2}+p_{010}^{2}+p_{011}^{2} \\
p_{100}^{2}+p_{101}^{2}+p_{110}^{2}+p_{111}^{2} \\
2(p_{001}p_{010}+p_{000}p_{011}) \\
2(p_{101}p_{110}+p_{100}p_{111}) \\
2(p_{000}p_{010}+p_{001}p_{011}) \\
2(p_{100}p_{110}+p_{101}p_{111}) \\
2(p_{000}p_{001}+p_{010}p_{011}) \\
2(p_{100}p_{101}+p_{110}p_{111})
\end{array}
\right) /P_{pass},
\end{equation}
where
$P_{pass}=(p_{000}+p_{001}+p_{010}+p_{011})^{2}+(p_{100}+p_{101}+p_{110}+p_{111})^{2}$
is the probability for the survived subset.

We argue that this subprotocol can be converted to a
prepare-and-measure subprotocol for secret sharing. In the
converted subprotocol, if $M_{A}\oplus M_{B}\oplus M_{C}=0 \mod
2$, the three parties keep the first trio. Otherwise, they
discard both trios.

The same conversion idea to a prepare-and-measure protocol also
applies to a concatenated protocol involving B' steps, P' steps
and random hashing. Therefore, to study the security of a secret
sharing protocol, we can go back to the GHZ distillation picture
and simply study the convergence of a GHZ distillation protocol
involving B', P' and random hashing. In what follows, we consider
only Werner-like states. By direct numerical calculation, we can
verify that our scheme used in secret sharing in a noisy channel
can distill GHZ states with nonzero yield whenever $F\geq 0.5372$
by some state-dependent sequence of B and P steps and then change
to our random hashing method if it works. For the optimal
sequences within 5 steps, we find it is (B'B'B'B'B') that just
gives $F\geq 0.5372$ followed by hashing method. Compared with
QKD scheme in a noisy channel, this lead to more entanglement
requirement for secret sharing than the QKD scheme ($F\geq
0.3976$) by using our decoding scheme. Similar to the case of
conference key agreement, the optimal sequences is for Werner
state with initial fidelity around $0.5372$. Different initial
fidelity requires different optimal sequence plus immediately
random hashing. After estimate reliably the shared GHZ-basis
diagonal states calculate an optimal sequence, the three parties
execute B' step by CCCC.  Whenever there is a P' step, they take
the parity $(X_{A}+X_{B}+X_{C})\mod 2$ of their own three
particles locally. No classical communication is needed. While we
have only applied our protocol to Werner-like states,
non-Werner-like states can be analyzed in a similar manner.

As in the case of conference key agreement, our protocol for
three-party secret sharing can be implemented by using only
bi-partite entanglement. The conversion is done by Alice
pre-measuring her subsystem. As a result, she only needs to
prepare some bi-partite entangled states and send them to Alice
and Bob.

As noted in the Introduction, in the above discussion, we have
assumed that the three parties are honest in executing the
protocol. While this may be a good assumption for conference key
agreement, it may not be a good assumption for quantum sharing of
classical secrets. In the latter protocol, whereas Alice may well
be honest, the participants, Bob and Charlie, have incentives to
cheat. We have not analyzed the question of how to make Bob and
Charlie honest in detail. This is an important question that
deserves future investigations. On the other hand, even in the
case of classical secret sharing, one might imagine that Bob can
cheat by claiming to hold a value that is different from what he
actually holds. We are not aware of any simple solution even in
this classical setting.

\subsection{Third-Man Cryptography}
\noindent
\label{ss:thirdman} We remark that a quantum sharing of classical
secrets protocol can also be used to implement third-man
cryptography. The goal of third-man cryptography is to allow a
party, say Alice, as a central server, to control whether two
other parties, Bob and Charles, can achieve secure key
distribution. If Alice refuses to co-operate, Bob and Charles
will fail to generate a secure key. On the other hand, if Alice
says that she is willing to help, then Bob and Charles can verify
that indeed they have managed generate a secure key, which is
secret even from Alice! Starting from our conference key
agreement protocol, Alice can simply broadcast all her
measurement outcome $Z_A$. With the knowledge of $Z_A$, Bob and
Charles should share some EPR states with each other. They can,
therefore, proceed to check the purity of their EPR pairs in a
similar manner as in standard BB84 and then proceed with
classical post-processing (advantage distillation, error
correction, and privacy amplification).

\subsection{Experimental Implementations}
\noindent
Previous experimental implementations of both quantum sharing of
classical secrets and third-man cryptography \cite{expQSCS}
required the simultaneous generation of two EPR pairs, which is
extremely challenging, whereas our protocols, which assume that
the preparer of the state is the central server, Alice, herself
have the distinctive advantage of requiring the generation of
only one EPR pair at a time. Our protocols, therefore, provide
substantial simplifications and much higher yield, compared to
previous experiment \cite{expQSCS}.

\subsection{Application of multi-partite entanglement}
\noindent
It will be interesting to study secret sharing with a more
general access structure \cite{gottesman00,smith00}. From an
application standpoint, it will be particularly interesting to
look into the possibility of converting those protocols to
``prepare-and-measure'' protocols where each party simply
receives some noisy version of an entangled state from a source
and then performs a measurement, followed by CCCCs.

\section{Significance of our protocols}
\noindent
\subsection{Protocols with two-way classical communications are
provably better than protocols with only one-way classical
communications}
\noindent
In, for example, our three-party conference key agreement
protocol, we allow Alice, Bob and Charlie to communicate back and
forth with each other. By doing so, they can achieve secure
conference key agreement for a Werner-like state whenever the
fidelity $F \geq 0.3976$. One might wonder whether a two-way
communication channel is generally needed between Alice and
others during the conference key agreement stage. In other words,
suppose Alice only has a broadcast channel to Bob and Charlie,
but is not allowed to receive any classical communications in the
conference key agreement stage. Starting with a Werner-like state
of the same fidelity, can Alice, Bob and Charlie still distill a
secure conference key?

As a comparison, for the BB84 quantum key distribution scheme,
which is a {\it two}-party prepare-and-measure common
key-agreement protocol, it is provably that protocols with
two-way classical communications are better than those without.
Indeed, in \cite{GLIEEE03}, BB84 with two-way classical
communications is proven to be secure up to an error rate of 18.9
percent whereas any one-way protocol will be insecure whenever
the bit error rate is larger than about 15 percent.

Returning to the {\it three}-party conference key agreement
scheme, is the best one-way protocol as good as a two-way
protocol? We now show directly that the answer is negative. In
other words, we will prove that three-party conference key
agreement protocols with two-way classical communications are
provably better than those without. We do so by demonstrating
that the tripartite Werner-like state (\ref{Werner}) with
fidelity $F=6/19$ cannot be distilled into GHZs by any one-way
protocols, but can be distilled by two-way protocols. Our
argument is similar to one given in Ref.~\cite{BDSW}.

One way for a preparer Trent to prepare a Werner-like state
(\ref{Werner}) with fidelity $F=6/19$ for three parties, Alice,
Bob and Charlie is the following. Suppose a preparer, say Trent,
prepares an ensemble of $n$ GHZ states. Each GHZ state consists
of a trio of qubits. For each trio, Trent gives one qubit to
Alice (A). What happen to the two remaining qubits of the trio?
Well, with a probability one half, he gives the two remaining
qubits to Bob (B) and Charlie (C) and a maximally mixed state to
David (D) and Eva (E). And, with a probability one half, he gives
the two remaining qubits to David (D) and Eva (E) instead and
gives a maximally mixed state to Bob (B) and Charlie (C). Now
suppose Alice, Bob and Charlie are interested in GHZ
distillation, but the three parties do not know which trios are
shared between Alice, Bob and Charlie, and which between Alice,
David and Eva. Then, the state of the trio can be described by a
density matrix of a tripartite Werner-like state (\ref{Werner})
with $\alpha=1/2$, i.e., $\rho _{W}= 1/2 \left\vert \Phi
^{+}\right\rangle \langle \Phi ^{+}|+\frac{1}{16}I$ and thus has
a fidelity $F=9/16$.

We now argue that such a Werner-like state with $F= 9/16$ cannot
be distilled without Alice receiving communications from Bob and
Charlie. This is done by a symmetry argument. Suppose there
exists a conference key agreement protocol that does not require
Alice to receive any communications from Bob and Charlie. Then,
in the end, Alice, Bob and Charlie will have a secret key, $k$,
safe from eavesdroppers. However, since the Bob/Charlie pair is
symmetric under interchange with the David/Eva pair, we can argue
that Alice, David, Eva must share the same secret key, $k$. But,
this contradicts the requirement that the key is secure against
any eavesdropper (because, clearly, Bob and Charlie know the key,
$k$).

In contrast, in Subsection~\ref{ss:reduction1}, we showed that
three-party secure conference key agreement can be achieved for a
Werner-like state with fidelity $F \geq 0.3976$, if two-way
classical communications are used. Therefore, any Werner-like
states with a fidelity $ 0.3976 \leq F \leq 9/16$ can be used for
secure conference key agreement with two-way protocols, but not
with any one-way protocols. This is a demonstration of the power
of two-way classical communications in multi-party quantum
cryptography.

\subsection{Comparison with violation of Bell inequalities}
\noindent
It is claimed in Refs.~\cite{Gisin1,Gisin2} that a violation of
Bell inequalities is a criterion for security of secret sharing
schemes \cite{hbb99} with the assumption that Eve would be able
to make only individual attacks. The security of protocols given
in Refs.~\cite{Gisin1,Gisin2} makes use of a {\it one}-way
protocol to extract a secret key. We remark that such a claim
does not apply to the present context where we allow the parties
to perform {\it two}-way communications. Violation of Bell
inequalities for $N$ particles Werner-like state is shown in
Ref.~\cite{zb2002} to be $\alpha >1/\sqrt{2^{(N-1)}}$. For a
tripartite system, this gives $\alpha >1/2$ and thus $F>9/16
\doteq 0.5625$. This is clearly a higher requirement for the
initial fidelity of a Werner-like state than that for our two-way
prepare and measure secret sharing scheme which only requires
$F\geq 0.5372$. Thus our two-way protocols are secure even when
Bell inequalities are not violated. This is another demonstration
of the power of two-way classical communications in multi-party
quantum cryptography.

\section{Concluding Remarks}
\noindent
In this paper, we study three multi-party quantum cryptographic
protocols: (a) conference key agreement, (b) quantum sharing of
classical secrets and (c) third-man cryptography. We start with a
protocol for GHZ distillation and convert it to a
``prepare-and-measure'' protocol for quantum cryptography. The
main requirement for a protocol to be convertible to a
prepare-and-measure protocol is that it cannot include any phase
error detection steps. This is because phase error patterns are
generally unavailable to the parties.

We remark that our three-party quantum cryptographic protocol can
be implemented by with the generation of a single entangled pair
of photons at a time. This is a major improvement over previous
experimental demonstrations \cite{expQSCS}.

Our protocol can be readily generalized to the case of $N >3$
parties and be implemented by using only $(N-1)$-partite
entangled states. In the course of our investigation, we
construct more efficient hashing protocols for GHZ distillation.
Our protocol applies to not only a GHZ state, but also a general
CSS-state. Moreover, we show that CSS-states are mathematically
equivalent to two-colorable graph states, thus putting the prior
work on distillation of two-colorable graph states
\cite{graph,graph1} in a more systematic setting.

We note that our protocols, which involves two-way classical
communications, are provably better than any protocol involving
only one-way classical communication. Furthermore, our protocols
work even when the initial state fails to violate the standard
Bell inequalities for GHZ-states.

However, we remark that our protocols are not proven to be
optimal. In future, it will be interesting to search for
protocols that can distill even noisier initial states. Except
for the case of hashing protocols, in this paper we have not
looked into the issue of yield closely. This can be an important
subject for future investigations.

It will also be interesting to a) explore further the connection
of our work with aforementioned subjects (e.g. non-two-colorable
graph state distillation) and b) generalize our results to
quantum sharing of classical secrets for more general access
structures. In the long term, it is our hope that such
investigations will shed some light on the fundamental questions
of the classification of multi-party entanglement and multi-party
entanglement distillation. In conclusion, the power and
limitations of multi-party quantum cryptography deserve future
investigations.

\section*{Acknowledgements}
\noindent
We thank helpful discussions with various colleagues including
Panos Aliferis, Hans Briegel, Daniel Gottesman, Peter Knight,
Manny Knill, Masato Koashi, Debbie Leung, Norbert L\"{u}tkenhaus,
John Preskill and Yaoyun Shi. We are indebted to Eric Rains for
showing that CSS states are two-colorable graph states. (Part~B
of Claim~2.) We also thank Xiongfeng Ma for some help in computer
programming. Much appreciation is indebted to the authors of
\cite{mppvk98,smolincm2002} for kindly approvement of using
figures in their papers. This work was supported in part by
Canadian NSERC, Canada Research Chairs Program, Connaught Fund,
Canadian Foundation for Innovation, Ontario Innovation Trust,
Premier's Research Excellence Award, Canadian Institute for
Photonics Innovations, University of Toronto start-up grant, and
the National Science Foundation under grant EIA-0086038 through
the Institute for Quantum Information at the California Institute
of Technology.

\end{document}